\documentclass[conference]{IEEEtran}
\IEEEoverridecommandlockouts
\usepackage{cite}
\usepackage{float}
\usepackage{amsmath,amssymb,amsfonts}
\usepackage{algorithmic}
\usepackage{graphicx}
\usepackage{textcomp}
\usepackage{xcolor}
\usepackage[hidelinks]{hyperref}
\usepackage{multirow}
\usepackage{booktabs}
\def\BibTeX{{\rm B\kern-.05em{\sc i\kern-.025em b}\kern-.08em
    T\kern-.1667em\lower.7ex\hbox{E}\kern-.125emX}}

\makeatletter
\def\ps@IEEEtitlepagestyle{%
  \def\@oddhead{}%
  \def\@evenhead{}%
  \def\@oddfoot{\parbox{\textwidth}{\scriptsize \copyright 2025 IEEE. Personal use of this material is permitted. Permission from IEEE must be obtained for all other uses, in any current or future media, including reprinting/republishing this material for advertising or promotional purposes, creating new collective works, for resale or redistribution to servers or lists, or reuse of any copyrighted component of this work in other works. Cite this article as follows: M. N. Hasan Fahim, M. A. Ullah Muhib, R. A. Tanvir, and A. Al Noman, ``Real-Time Earthquake Magnitude Classification from Initial P-Waves: Models, Dataset, and Comparative Analysis for South Asia,'' \textit{2025 28th International Conference on Computer and Information Technology (ICCIT)}, Cox's Bazar, Bangladesh, 2025. DOI: \href{https://ieeexplore.ieee.org/document/11489542}{10.1109/ICCIT68739.2025.11489542.}}}%
  \def\@evenfoot{}%
}
\makeatother

\begin{document}

\title{Real-Time Earthquake Magnitude Classification from Initial P-Waves: Models, Dataset, and Comparative Analysis for South Asia}

\author{
    \IEEEauthorblockN{
        Md Nasiat Hasan Fahim\IEEEauthorrefmark{1},
        Md. Abid Ullah Muhib\IEEEauthorrefmark{2},
        Rayhanul Amin Tanvir\IEEEauthorrefmark{3},
        Abdullah Al Noman\IEEEauthorrefmark{4}
    }
    
    \IEEEauthorblockA{
        Department of Computer Science and Engineering \\
        Shahjalal University of Science and Technology \\
        Sylhet, Bangladesh \\
        Emails: 
        \IEEEauthorrefmark{1}nhfahim18@gmail.com,
        \IEEEauthorrefmark{2}uusshas12@gmail.com,
        \IEEEauthorrefmark{3}rayhanulamint2@gmail.com,
        \IEEEauthorrefmark{4}nomancseku@gmail.com
    }
}

\maketitle

\begin{abstract}
Rapid earthquake magnitude estimation is crucial for effective early warning systems that can save lives and reduce economic damage. In this paper, we present a comprehensive study of magnitude classification using only the vertical component of the initial 7-second P-wave window from a single station. We compare six machine learning approaches that range from traditional models to state--of-the-art deep learning architectures. We also curated a novel dataset of 7,318 earthquake events in South Asia. The dataset was categorized into five Richter-scale classes: slight (3.0--3.9), light (4.0--4.9), moderate (5.0--5.9), strong (6.0--6.9) and severe ($\geq$\,7.0). Our experiments show that deep learning models substantially outperform traditional approaches. Our Transformer based architecture achieved 76.23\% standard accuracy and 81.56\% adaptive accuracy with 4.8 ms inference latency. The adaptive-accuracy metric is introduced for the inherent uncertainty in magnitude estimation of near class boundaries. These results indicate that the attention mechanisms in Transformers combined with adaptive classification effectively capture the temporal dynamics of seismic signals. The architectural advantage facilitates promising generalization to rare high-magnitude events, despite the inherent data scarcity characteristic of seismic catalogs. The adaptive accuracy provides a more realistic assessment of model performance, and the result suggests viability for real-time deployment.
\end{abstract}


\section{Introduction}
As earthquakes continue to pose catastrophic global risks, reliable Earthquake Early Warning (EEW) systems are vital. This is critical for South Asian urban agglomerates such as Dhaka, where seismic vulnerability is acute ~\cite{mishra2020seismic}. Traditional earthquake magnitude estimation methods require complete waveform data from multiple stations, resulting in frequent delays that limit their effectiveness for early warning applications.  

The trade-off between speed and accuracy is the fundamental challenge in EEW systems. Although for accurate magnitude estimation, more data is required, it loses the purpose of early warning. The fastest seismic waves with valuable information on the characteristics of an earthquake that arrives first at seismic stations are the P-waves~\cite{11}. Recent studies show that the initial P-wave can provide sufficient information for magnitude estimation~\cite{Nazeri2017}.

In this work, we present a comprehensive and comparative study of machine learning approaches for earthquake magnitude classification using only the initial 7-second P-wave window. We also curated a novel P-wave dataset of South Asia for standardized benchmarking of this region. Our key contributions include

\begin{itemize}
\item A systematic evaluation of six different machine learning models, from traditional methods to advanced transformer architectures, on a consistent dataset and standardized evaluation framework.
\item A standardized benchmark dataset of 7,318 quality-controlled South Asian single-station vertical component earthquake events to facilitate reproducible research in a region with complex tectonic settings.
\item Introduction of an 'Adaptive Accuracy' metric, designed to quantify model performance near magnitude class boundaries where inherent estimation uncertainty is high.
\item Demonstration of the operational feasibility of single-station magnitude estimation, achieving 81.56\% adaptive accuracy with an inference latency of just 4.8 ms.
\end{itemize}

\section{Related Work}

Traditional earthquake magnitude estimation techniques use early P-waves to predict earthquake sizes. Following Richter’s instrumental scale\cite{22}, on-site methods typically extract features such as peak displacement ($P_d$), predominant period ($\tau_p$) and average period ($\tau_c$) from the initial few seconds of the signal\cite{5,6}. 
Although the $P_d$ approach underpins many operational early-warning systems\cite{7}, it suffers from high uncertainty and begins to saturate for strong events. 
Combining several parameters can reduce error, but these multi-parameter methods still rely on hand-crafted features that may overlook the informative signal structure\cite{8}.

The first machine-learning studies in seismology employed algorithms such as support-vector machines and random forests\cite{9}. 
These methods improved detection and classification compared to rule-based schemes, yet they still relied on manually engineered features. 
Deep learning removed that bottleneck: convolutional neural networks (CNNs) can learn features directly from raw waveforms and have already advanced phase picking, event detection, and magnitude estimation in both global and regional networks \cite{10,11,12,WIBOWO2024100194}.

Within deep learning, single-station CNNs now deliver magnitude estimates that rival traditional multi-station techniques, and fully convolutional designs and lightweight attention networks allow real-time operation\cite{guo2025lftnet,14}. 
Sequence models that exploit temporal evolution—long short-term memory networks (LSTMs) and their bidirectional variants—capture longer-range dependencies and further boost performance\cite{16}. Transformer networks, which are dependent on self-attention, have been adapted to seismic data; by focusing on the most informative parts of the waveform, they achieve even higher accuracy\cite{17,18}.
Recent work has further validated this approach: Zhang et al.\cite{zhang2024realtime} demonstrated that transformers effectively fuse multimodal seismic data for real-time magnitude estimation, while Münchmeyer et al.\cite{munchmeyer2021earthquake} showed that attention-based models can generalize better to rare, high-magnitude events crucial for early warning and wavefield forecasting \cite{lyu2024wavecastnet}.

Despite this progress, several challenges remain. Training data suffers from significant imbalance due to the lack of large magnitude events. Variation in crustal and attenuation properties leads models to performance degradation in different tectonic settings. Data acquisition, processing and decision making must be done within a few seconds since Early-warning systems impose strict latency constraints. Finally, current models lack estimates of their own uncertainty, which is essential to make operational decisions. The present study addresses these limitations by benchmarking diverse architectures on a unified South-Asian dataset using a novel adaptive accuracy metric.

\section{Dataset and Preprocessing}
Our dataset contains earthquake events from the South Asian region, an area characterized by complex tectonic settings due to the collision between the Indian and Eurasian plates. These waveform data were collected directly from multiple seismic networks of the IRIS Data Management Center (IRIS~DMC) over a 30-year period (1995-2025), forming a curated and standardized dataset for P-wave magnitude classification with comprehensive coverage of magnitude ranges and epicentral distances.

The final dataset consists of 7,318 high-quality earthquake events after rigorous quality control and preprocessing from the initial dataset of 18,325 records. Each event is formed by: 1) One-component acceleration waveforms. 2) Precise P-wave and S-wave arrivals. 3) Validated earthquake magnitude from multiple agencies. 4) Information about the epicentral distance and the depth. 5) Location, site conditions and other station metadata.

\begin{figure}[ht]
\centering
\includegraphics[width=\columnwidth]{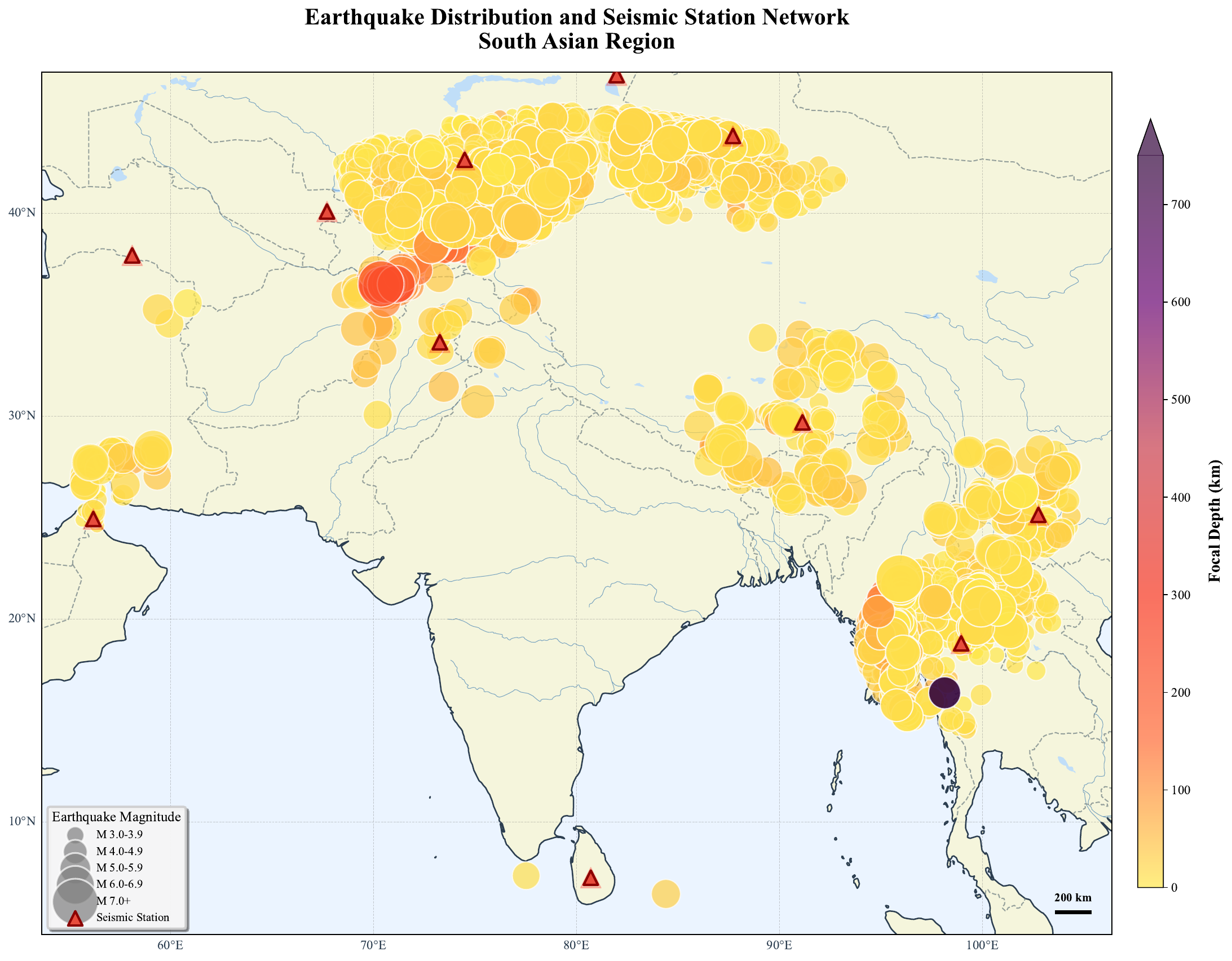}
\caption{Earthquake and seismic station network distribution in the South Asian region. Circle sizes represent earthquake magnitudes and colors indicating focal depth. Red triangles mark seismic station locations used for data collection.}
\label{fig:earthquake_map}
\end{figure}

Figure \ref{fig:earthquake_map} illustrates the spatial distribution of earthquakes in the dataset. The dense cluster in the Himalayan arc represents the frequent collision between the Indian and the Eurasian plates. However, the peninsular Indian zone represents the less frequent events.

Table \ref{table:earthquake_margnitude_classification} shows the categorized earthquake events into five classes based on their potential impact and follows standard seismological practice and EEW system requirements.

\begin{table}[ht]
\centering
\caption{Earthquake Magnitude Classification}
\label{table:earthquake_margnitude_classification}
\begin{tabular}{llcc}
\toprule
Class & Category & Magnitude Range & Impact Level \\
\midrule
0 & Slight & 3.0 - 3.9 & Minimal \\
1 & Light & 4.0 - 4.9 & Minor \\
2 & Moderate & 5.0 - 5.9 & Moderate \\
3 & Strong & 6.0 - 6.9 & Severe \\
4 & Severe & $\geq$ 7.0 & Catastrophic \\
\bottomrule
\end{tabular}
\end{table}

To ensure data integrity, we applied a rigorous quality control protocol. 1) Removal of events where the S-wave arrives before the P-wave. 2) Signal-to-Noise Ratio (SNR) filtering by comparing the Root Mean Square (RMS) amplitudes of a 1.5-second signal window against a 3-second noise window for both phases. 3) Verification of phase picking accuracy using an automated Short-Term Average/Long-Term Average algorithm. 4) Elimination of non-physical artifacts by grouping events into 0.5-magnitude bins and removing outliers outside the amplitude percentiles. This filtering primarily targeted the M3.0–3.9 range, which is susceptible to instrument glitches in weaker signals.

To maintain waveform quality, we designed a preprocessing pipeline. DC offsets were removed by subtracting the mean of the 5-second pre-arrival window. A fourth-order Butterworth bandpass filter was used to suppress noise while preserving seismic characteristics. All waveforms were resampled at 15 Hz for consistency. Normalization was formed by applying standard Z-score scaling that uses statistics derived from the training set. Finally, the input features were generated by extracting a fixed 7-second window starting from the arrival time of the P-wave ($t_p$), resulting in a representative feature vector of 105 samples for input to the model.

\begin{figure}[ht]
\centering
\includegraphics[width=\columnwidth]{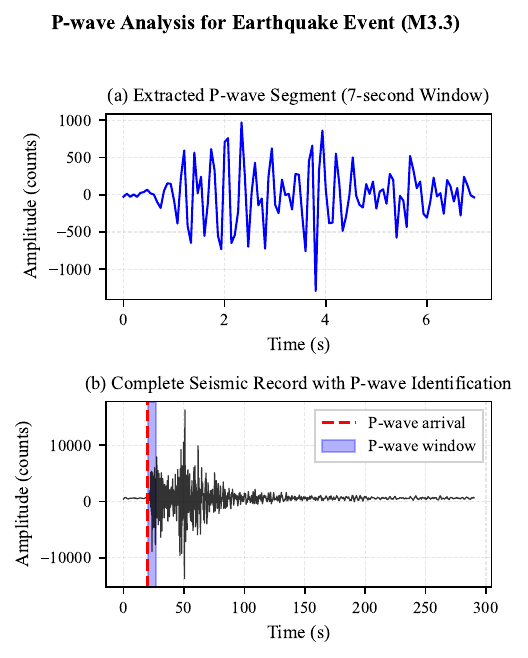}
\caption{P-wave analysis of a magnitude 3.3 earthquake event. (a) Extracted 7-second P-wave segment showing the characteristic waveform pattern used for classification. (b) Complete seismic record with P-wave arrival marked and the extraction window highlighted.}
\label{fig:pwave_analysis}
\end{figure}

Figure \ref{fig:pwave_analysis} demonstrates our P-wave extraction methodology. We selected a 7-second window to accommodate observed inconsistencies in arrival annotations, where actual P-wave onsets frequently lagged $>3$ seconds behind timestamps. This wider buffer ensures the signal is captured despite labeling noise.

\begin{figure}[ht]
\centering
\includegraphics[width=\columnwidth]{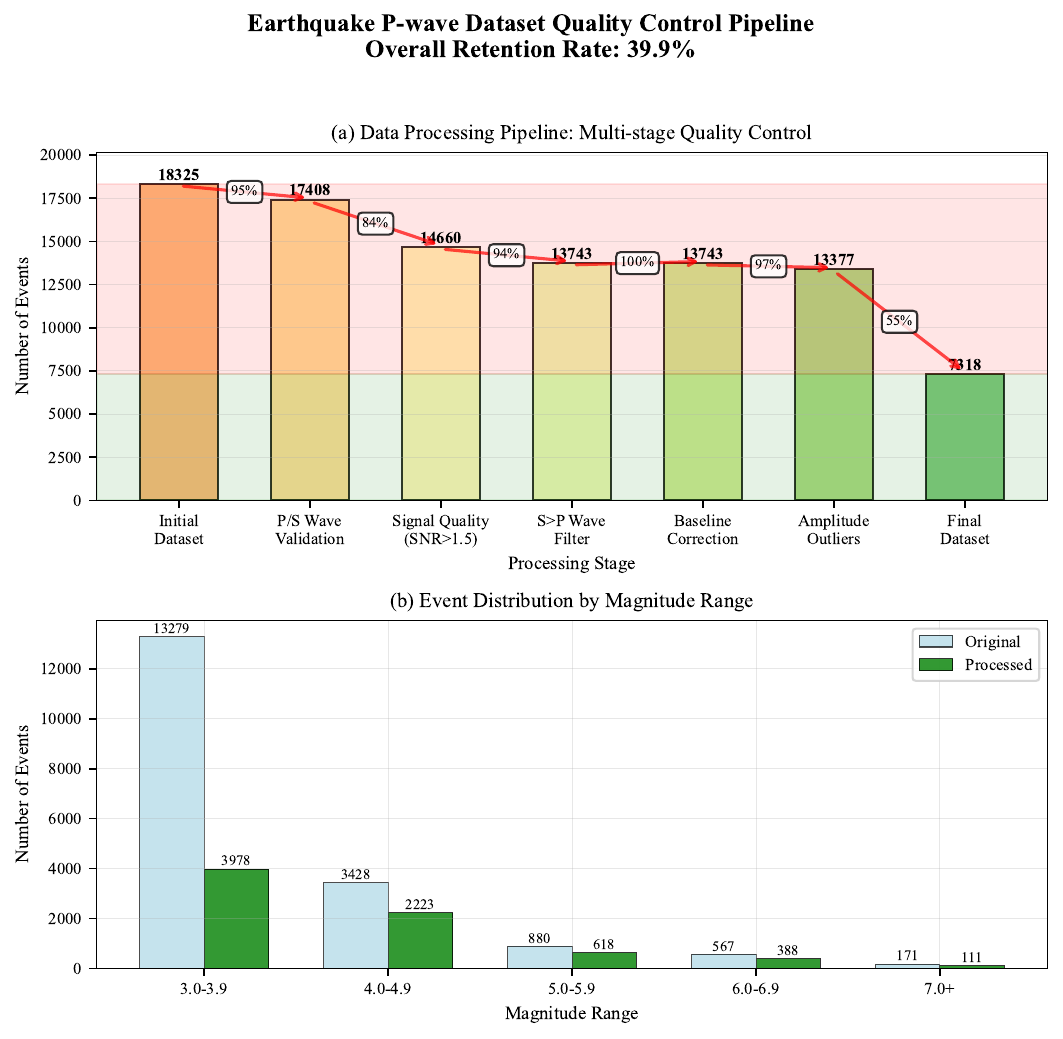}
\caption{Earthquake P-wave dataset quality control pipeline showing (a) multi-stage processing with retention rates at each step, and (b) event distribution by magnitude range before and after processing. The overall retention rate of 39.9\% reflects our stringent quality requirements.}
\label{fig:data_pipeline}
\end{figure}

As shown in Figure \ref{fig:data_pipeline}, the rigorous quality control pipeline reduced the dataset from 18,325 to 7,318 events. P/S wave validation (5\% loss) and the assessment of signal quality based on SNR thresholds (16\% loss) causes the most significant reductions. The final distribution shows the retention of a representative sample across all magnitude ranges and the preservation of the natural predominance of smaller earthquakes.

\begin{table}[ht]
\centering
\caption{Dataset Distribution}
\begin{tabular}{lrrr}
\toprule
Split & Samples & Percentage & Purpose \\
\midrule
Training & 6,124 & 83.7\% & Model training \\
Validation & 681 & 9.3\% & Hyperparameter tuning \\
Test & 513 & 7.0\% & Final evaluation \\
\bottomrule
\label{table:final_dataset}
\end{tabular}
\end{table}

The final dataset in table \ref{table:final_dataset} reveals a significant imbalance in class distribution, reflecting the natural frequency of earthquake magnitudes. This imbalance in the dataset poses challenges for model training, which we address through weighted sampling and specialized loss functions, as described in Section IV.

\section{Methodology}
We implemented and evaluated six models spanning traditional machine learning to state-of-the-art deep learning architectures, each processing 7-second P-wave windows (105 samples at 15 Hz).

1) Logistic Regression: Our baseline model performs multi-class classification using softmax on the flattened 105-dimensional P-wave input, with L2 regularization. This simple linear approach provides a performance floor for comparison.

2) XGBoost: We employed gradient boosting with 500 estimators, a maximum depth of 6, a learning rate of 0.05, and 80\% feature/sample subsampling. Features include the raw waveform of 105 points plus statistical descriptors (mean, standard deviation, RMS, min/max values).

3) CNN1D-Basic: The basic CNN architecture employed multi-scale convolutional blocks processing the input through parallel convolutions (kernels: 3, 5, 7, 9) to capture features at different temporal scales. The network consists of an initial multi-scale block producing 64 channels, followed by max pooling and 9 residual blocks progressively increasing channels (64→128→256→512→1024). The Global average pooling aggregates temporal features before a 3-layer classifier (1024→512→256→128→5) with batch normalization and dropout (0.5, 0.3, 0.2) at each layer.

4) CNN1D-ResNet: Our ResNet variant implements a deeper architecture with enhanced residual connections. After an initial multi-scale stem (1→64 channels), the network employs 4 residual blocks with strided convolutions for downsampling (64→128→256→512→512). Each residual block contains two 3×1 convolutions with batch normalization and skip connections. Multi-scale feature extraction at the output (using 1×1, 3×1, and 5×1 convolutions) creates a 384-dimensional representation, followed by a 3-layer classifier with progressive dropout.

5) BiLSTM: The bidirectional LSTM processes sequences in both temporal directions using 4 stacked layers with 512 hidden units each and 0.3 dropout between layers. The architecture captures long-range dependencies by maintaining separate forward and backward hidden states, which are concatenated and averaged across time steps. The resulting 1024-dimensional representation passes through a 4-layer classifier (1024→512→256→128→5) with batch normalization and decreasing dropout rates.

6) Transformer: We propose a Deep Attention Transformer framework designed specifically to model the non-linear temporal dynamics of seismic P-waves. In contrast to the sequential processing of BiLSTMs or the local feature extraction of CNNs, our proposed attention-based framework utilizes a self-attention mechanism to capture global dependencies across the entire 7-second window. The framework projects the input to 512 dimensions and adds learnable positional encodings to preserve temporal order without relying on recurrence. Eight encoder layers, each with 16-head self-attention and 1024-dimensional feed-forward networks, process the sequence. We employ attention-based pooling where learned attention weights aggregate temporal features into a global representation. The classifier mirrors the BiLSTM design with 4 layers and progressive regularization. All weights are initialized using the Xavier uniform distribution.

To mitigate the severe class imbalance resulting from the Gutenberg-Richter law (power-law distribution of earthquake sizes), we employ weighted random sampling using smoothed inverse class frequencies during training, ensuring balanced mini-batches, and Focal Loss with $\alpha$=1.0 and $\gamma$=2.0 for deep models, which down-weights easy examples and focuses learning on hard, misclassified instances. We apply online augmentation with 75\% probability during training, including Gaussian noise addition ($\sigma$=0.02), random time shifts (up to 15\% of sequence length), amplitude scaling (±20\%) and frequency-domain phase perturbations to improve generalization without altering fundamental seismic characteristics. All deep learning models use AdamW optimizer with learning rate $10^{-4}$ and weight decay $10^{-4}$, implementing linear warm-up over 5 epochs followed by linear decay, gradient clipping (max norm 1.0) to prevent training instability, and early stopping with patience of 10 epochs based on validation accuracy to prevent overfitting.

All models were implemented in PyTorch 2.0 and trained on NVIDIA P100 GPU using mixed precision training for efficiency, with consistent data preprocessing across all models including DC offset removal, bandpass filtering (1-20 Hz), normalization using training dataset statistics. The models were trained for a maximum of 200 epochs (CNNs), 100 epochs (BiLSTM) or 50 epochs (Transformer) with batch size 32 for deep models and 64 for logistic regression, using a stratified 83.7\%/9.3\%/7.0\% train/validation/test split by magnitude class and seed 42 for reproducibility. Training times on a single 16GB VRAM P100 GPU ranged from 0.1 hours (logistic regression) to 6 hours (Transformer), with inference requiring 0.1-4.8 ms per sample to meet real-time constraints, while model sizes vary from 530 parameters (logistic regression) to 8.7M parameters (Transformer), with the BiLSTM (3.2M) offering a good accuracy-efficiency trade-off.

We evaluate models using standard multi-class accuracy and introduce an adaptive accuracy metric that tolerates one-class deviations for samples within ±0.2 magnitude units of class boundaries (3.0, 4.0, 5.0, 6.0, 7.0), reflecting the inherent uncertainty in magnitude estimation near these boundaries and providing a more realistic assessment for operational deployment.

\section{Results and Analysis}
The results in table \ref{table:model_performance_comparison} demonstrate a clear progress in performance from traditional methods to deep learning architectures. The Transformer model achieves the best performance with 76.23\% standard accuracy and 81.56\% adaptive accuracy, representing a 144\% improvement over the baseline logistic regression model.

\begin{table}[ht]
\centering
\caption{Model Performance Comparison}
\begin{tabular}{lcccc}
\toprule
Model & Accuracy & Adaptive & Parameters \\
      & (\%) & Accuracy (\%) \\
\midrule
Logistic Reg. & 31.23 & 37.23 & 530 \\
XGBoost & 43.22 & 48.43 & 500K \\
CNN1D-Basic & 47.08 & 52.79 & 2.8M \\
CNN1D-ResNet & 55.76 & 61.32 & 5.4M \\
BiLSTM & 65.79 & 72.19 & 3.2M \\
Transformer & \textbf{76.23} & \textbf{81.56} & 8.7M \\
\bottomrule
\label{table:model_performance_comparison}
\end{tabular}
\end{table}

\begin{figure}[ht]
\centering
\includegraphics[width=\columnwidth]{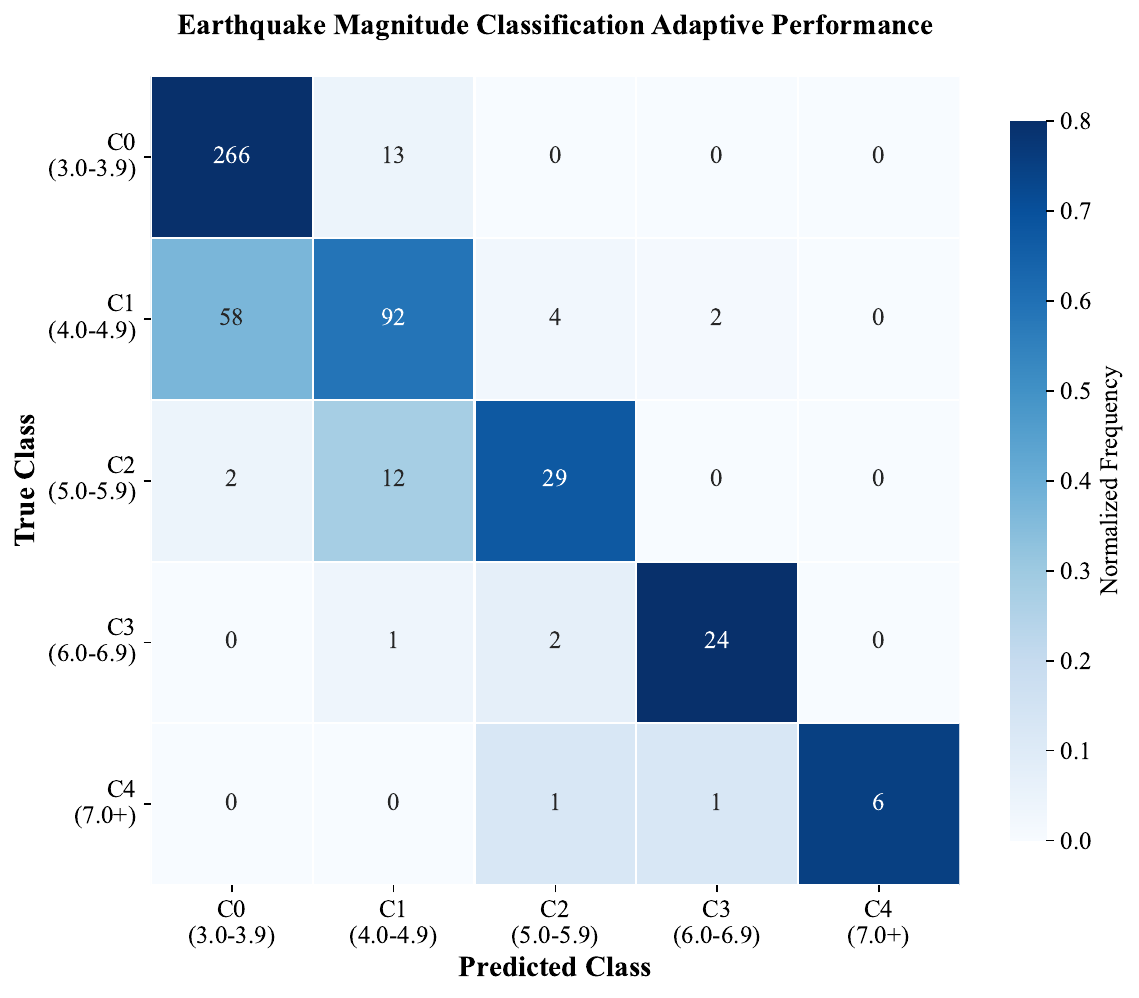}
\caption{Confusion matrix showing adaptive performance of the Transformer model on test data. Values indicate the number of samples classified into each category, with darker colors representing higher frequencies. The adaptive metric allows one-class deviation for samples near magnitude boundaries.}
\label{fig:confusion_adaptive}
\end{figure}

\begin{figure}[ht]
\centering
\includegraphics[width=\columnwidth]{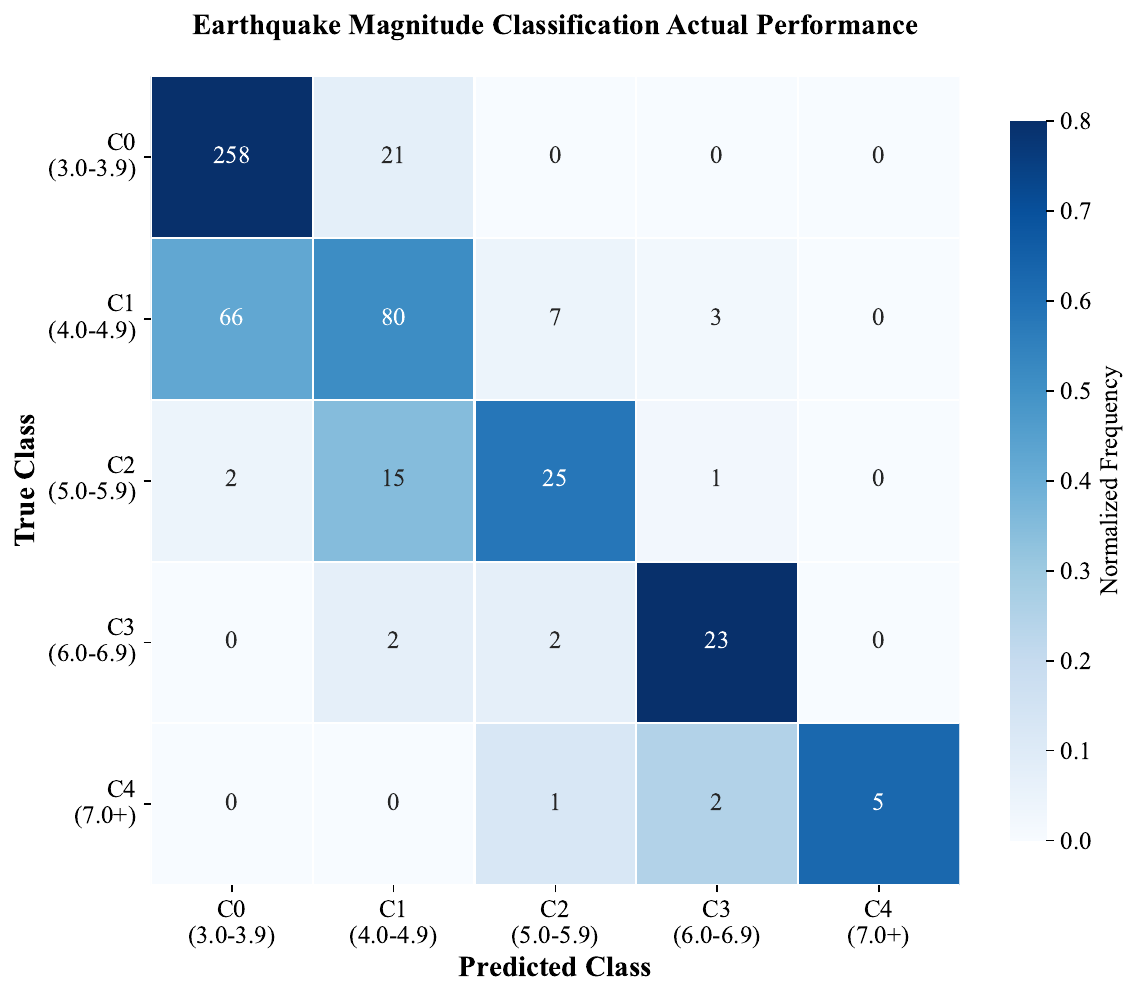}
\caption{Confusion matrix showing actual (standard) performance of the Transformer model on test data. This stricter metric requires exact class matches without boundary tolerance.}
\label{fig:confusion_actual}
\end{figure}

The confusion matrices in figures \ref{fig:confusion_adaptive} and \ref{fig:confusion_actual} reveal diagonal dominance, where the adaptive metric yields 266/279 (95.3\%) correct for Class 0 and 92/156 (59.0\%) for Class 1. Errors concentrate between adjacent classes (e.g., 57 Class 1 predicted as Class 0 near M4.0), are biased toward underestimation, and worsen for rare high-magnitude events (Class 4: 6/8 adaptive, 5/8 actual), reflecting class imbalance.

The relationship between model complexity and performance reveals important information for practical implementation. Although the Transformer model with  achieves the highest accuracy of (76.23\%), it has a large parameter of 8.7M, whereas the BiLSTM achieves 65.79\% accuracy with only 3.2M parameters. This performance gap demonstrates efficient parameter utilization. It also suggests that architectural choices are more significant than the raw parameter count for seismic data processing.

The diminishing returns observed with increasing complexity indicate that future work may require novel architectures rather than simply scaling existing ones. The BiLSTM's strong performance relative to its size makes it a practical option for resource-constrained deployment scenarios.

We examined 121 misclassified samples (23.77\%). 73 (60.3\%) samples from these misclassifications lie within 0.2 magnitude units of class boundaries (31 near 4.0; 24 near 5.0; 15 near 6.0; 3 near 7.0), supporting the adaptive metric's tolerance.

Table \ref{table:computational_requirements} represents the computational requirements of the models. All models were trained under the same settings with sub-5 ms GPU inference, meeting real-time needs; deployment should weigh accuracy versus cost.

\begin{table}[ht]
\centering
\caption{Computational Requirements}
\begin{tabular}{lccc}
\toprule
Model & Training & Inference & Memory \\
      & Time (h) & Time (ms) & (MB) \\
\midrule
Logistic Reg. & 0.1 & 0.1 & 0.1 \\
XGBoost & 0.5 & 0.5 & 2.0 \\
CNN1D-Basic & 2.0 & 1.2 & 11.2 \\
CNN1D-ResNet & 3.5 & 2.1 & 21.6 \\
BiLSTM & 4.0 & 3.5 & 12.8 \\
Transformer & 6.0 & 4.8 & 34.8 \\
\bottomrule
\label{table:computational_requirements}
\end{tabular}
\end{table}

\section{Discussion}
Our evaluation demonstrates that deep learning models substantially outperform traditional methods, which underscores the value of learned temporal representations. The performance hierarchy observed in our results validates the proposed attention-based framework. Although CNNs effectively extract local frequency features, they lack the capacity to model long-range temporal dependencies. Similarly, BiLSTM introduces sequential memory, but it is structurally limited by the information bottleneck inherent in recurrence. Due to the limitation, BiLSTM struggles to preserve dependencies between the P-wave onset and later signal phases in the 105-sample window. Our proposed Transformer framework overcomes these limitations through its attention mechanism. This mechanism allows the transformer to compute the relevance of every time step relative to every other time step ($N \times N$ complexity). Context-aware processing in the transformer isolates critical seismic phase arrivals regardless of their position in the window. This global context modeling drives the 10.44\% accuracy gain over the strongest baseline (static $<$ local CNN $<$ sequential BiLSTM $<$ global Transformer), with a 10.44\% absolute gain from BiLSTM to Transformer. The adaptive metric consistently adds 5–6\% across models, indicating systematic boundary uncertainty. Per-class F1 for the Transformer ranges 0.68–0.89, with strong performance at extremes despite the imbalance.

Using only 7-second P-wave windows, our approach can deliver magnitude estimates 15–25 seconds earlier than methods awaiting S-waves (e.g., 21s vs. 40s at 100 km; net gain 19s), with 4.8 ms inference enabling real-time use. Single-station processing reduces infrastructure and network dependence, supports edge deployment (34.8 MB), and scales linearly. Confusion-derived probabilities enable risk-based thresholds. Our curated South Asian dataset collected from IRIS DMC facilitates reproducible evaluation and region-specific operational tuning, and operationalization benefits from local fine-tuning, continuous learning, and modest ensembles for robustness.

Cross-regional tests show degradation, indicating the need for domain adaptation and region-specific fine-tuning. For M$\geq$7.0 (n=8), underestimation appears for the largest events (e.g., M7.8 misclassified as Class 3), suggesting 7-second windows may not capture full rupture. Operational constraints include $\geq$15 Hz sampling, accurate P picks (±0.5 s), handling simultaneous events, and real-time QC. Future iterations will investigate the latency-accuracy trade-off by evaluating shorter windows. We also aim to implement rigorous re-annotation of P-wave arrival times to eliminate the label noise necessitating the current wider buffer. Near-term work will add uncertainty quantification (e.g., MC Dropout, small ensembles targeting 90\% coverage), explore multi-task learning (magnitude/depth/azimuth), integrate physics-informed priors, and evaluate federated training for privacy-preserving adaptation.

\section{Conclusion}
This work establishes that accurate and rapid magnitude classification is feasible using only the initial 7-second P-wave window. Across our curated South Asian dataset collected directly from IRIS DMC, deep learning methods, particularly a Transformer architecture consistently outperform traditional baselines, achieving 76.23\% standard accuracy and 81.56\% adaptive accuracy with sub-5 ms inference. The adaptive accuracy metric offers a practically meaningful view of performance by acknowledging uncertainty near magnitude boundaries, and attention mechanisms provide an effective way to capture the evolving temporal dynamics in early arrivals.

Beyond model performance, this study also contributes in the construction of a novel and standardized P-wave dataset for the South Asian region. The dataset supports reproducible benchmarking and can serve as a foundation for further research in EEW via a curated pipeline (quality control, preprocessing, and consistent labeling). Our analysis reveals expected error modes (adjacent-class confusions near boundaries and underestimation for rare large events) and highlights that improved temporal modeling capacity primarily arises architectural.

Although this study suffers from cross-region generalization and the 7-second window may be insufficient for the most extreme events, these limitations suggest clear, actionable directions. Overcoming these gaps with the provided dataset and evaluation framework can lead to a robust, real-world deployment. In summary, the synergy of rigorous data curation and attention-based modeling establishes a pathway toward reliable, single-station magnitude estimation, offering a cost-effective solution for densifying EEW networks in seismically active developing regions.

\bibliographystyle{IEEEtran}
\bibliography{main}

\end{document}